\documentclass[final,3p,times,twocolumn]{elsarticle}
\usepackage{hyperref}[colorlinks=true, linkcolor=blue] 
\hypersetup{
    colorlinks = true,
    linkcolor=blue,
    citecolor=black,
    urlcolor=blue,
}

\usepackage{amssymb}
\usepackage{amsmath}

\begin{document}

\begin{frontmatter}




\title{Rethinking Total Absorption Gamma Spectroscopy Deconvolution: 
Supervised Machine Learning vs Response-Matrix Methods}

\author[label1]{J. Balibrea-Correa} 
\author[label1]{E. N{\'a}cher}
\author[label1]{C. Fonseca-Vargas}
\author[label1]{J. L. Tain}
\affiliation[label1]{organization={Instituto de F\'{\i}sica Corpuscular, CSIC - Universidad de Valencia, Spain},
            postcode={46980}, 
            state={Valencia},
            country={Spain}}

\begin{abstract}
The extraction of $\beta$-feeding distributions in Total Absorption $\gamma$-ray Spectroscopy constitutes a challenging inverse problem, particularly in nuclei with complex decay schemes involving a large number of excited states. In such cases, the measured spectrum arises from the superposition of many detector response functions, making the determination of the individual feedings intrinsically ill-posed and highly sensitive to the methodology employed.
In this work, we present a systematic comparison between supervised Machine-Learning techniques and Response-Matrix methods using realistic Monte Carlo simulations of an experimental Total Absorption Spectrometer. Supervised Machine-Learning approaches construct a non-parametric estimator that infers level feedings from the measured spectrum after a training stage, whereas Response-Matrix methods determine the feeding distribution by directly minimizing the difference between measured and reconstructed spectra. Our results show that supervised Machine-Learning techniques achieve superior accuracy in the reconstruction of individual feeding intensities, whereas Response-Matrix methods provide robust and physically consistent initial solutions. These findings support a hybrid strategy in which a Response-Matrix method is first used to obtain an initial feeding estimate, which is then refined using a supervised Machine-Learning approach to achieve improved overall accuracy.
\end{abstract}





\end{frontmatter}



\section{Introduction}\label{sec1}

The analysis of experimental data in Total Absorption $\gamma$-ray Spectroscopy (TAGS) requires solving complex inverse problems to remove the response function of the detector and recover the underlying physical quantities of interest~\cite{Tain:07}. This step, often called deconvolution, or “unfolding”, is essential for accurate interpretation of the physical data. Foundational studies that quantified and compared unfolding algorithms in $\gamma$-ray spectroscopy, particularly for reconstructing incident $\gamma$-ray spectra from scintillation counter data, explored approaches such as linear regularization~\cite{Press:92}, maximum-likelihood estimation via expectation-maximization~\cite{Dempster:77,Shepp:82} which is identical to the Bayesian iterative method~\cite{Richardson:72,Lucy:74,DAGOSTINI:95}, and the maximum-entropy method~\cite{Shore:80,Tikochinsky:84,Titterington:1985}. Meng and Ramsden~\cite{Meng:00} further demonstrated the feasibility of applying these advanced data-analysis techniques to standard NaI detectors, with an emphasis on applications where obtaining accurate estimates of counts in specific full-energy peaks is critical.

However, the challenge of unfolding TAGS data presents unique complexities compared to traditional high-resolution $\gamma$-spectroscopy (HRGS)~\cite{Tain:07,Tain:07a}. HRGS typically relies on detecting individual $\gamma$-rays to construct decay level schemes. In contrast, TAGS aims to determine directly the $\beta$-intensity distribution by detecting the entire de-exciting $\gamma$-ray cascade, using highly efficient, scintillation-based, 4$\pi$ experimental devices, and analyzing the full spectral shape rather than discrete peaks of the spectrum~\cite{DUKE:70,Rubio:17}. This method is essential for studying complex decays involving a large number of excited levels, which in HRGS are particularly prone to the systematic errors known as the \textit{Pandemonium} effect~\cite{Hardy:77}. In nuclei with large decay-energy windows and high level densities, $\beta$ decay can populate states at high excitation energies that de-excite through complex $\gamma$-ray cascades. The primary, high-energy $\gamma$-rays emitted in these cascades are frequently missed in HRGS measurements, since they often deposit only part of their energy before escaping the detector, contributing to the Compton continuum rather than to full-energy peaks. As a result, the $\gamma$-intensity balance, based on peak analysis and used to infer the $\beta$-feeding distribution, becomes systematically biased, leading to an artificial redistribution of $\beta$ feeding towards lower excitation energies and, consequently, to an underestimation of the feeding to highly excited states~\cite{Hardy:77}.

It is worth emphasizing that the energy resolution of Total Absorption Spectrometers (TAS) used in the TAGS technique is typically limited (of the order of 150~keV at 3~MeV). Consequently, the TAGS analysis does not focus on resolving narrow, discrete transitions of energy $E_{\gamma}$, as is the case in HRGS~\cite{Meng:00}, but rather on determining the average feeding probability over finite excitation energy ($E_{x}$) intervals and on characterizing the overall shape of the distribution.

In any TAGS analysis, the primary objective is to determine the $\beta$-intensity or normalized $\beta$-feeding distribution, $\vec{f}$, that is, the set of $\beta$-decay population probabilities to the excited states of the daughter nucleus. Each component $f_{j}$ represents the feeding probability to the individual level $j$ that exhibits a characteristic de-excitation pattern that can be observed with a TAS.

Due to finite experimental detection efficiency, energy resolution, incomplete $\gamma$-ray absorption and the interaction of $\beta$-decay particles (e$^{-}$/e$^{+}$, X rays, etc.), each decay pattern produces a characteristic spectral shape, commonly referred to as the detector response function. These response functions can not be determined directly from experimental data and must therefore be evaluated through detailed Monte Carlo (MC) simulations bench-marked on calibration measurements. A detailed description of the MC-based TAS response calculation performed in this work is provided in Sec.~\ref{Sec:Meth}.


Because the $\beta$ decay populates multiple excited states in the daughter nucleus, the experimentally measured TAS spectrum, $\vec{d}$, represents a superposition of the corresponding e$^{-}$ and $\gamma$-cascade response patterns rather than a direct mapping of the primary $\beta$-feeding intensities. Mathematically, the relationship between the observed TAS spectrum and the feeding distribution can be expressed as

\begin{equation}
    \begin{aligned}
\vec{d} & = N \mathbf{R}\cdot\vec{f}, \\
d_{i} & = N \sum_{j} R_{ij} f_{j}
\end{aligned}\label{eq:1}
\end{equation}

where $\mathbf{R}$ represents the response matrix and $N$ the number of decays.

The extraction of $\vec{f}$ therefore requires solving the inverse problem defined by Eq.~\ref{eq:1}. A direct inversion of the response matrix $\mathbf{R}$ is, however, numerically unstable and generally impracticable~\cite{Tain:07,Tain:07a}. This problem belongs to the class of ill-posed inverse problems, since the measured data are only weakly sensitive to variations of the feeding distribution. In addition, it is an ill-conditioned problem since the detector response to feeding of adjacent excitation-energy bins, especially at high excitation energies, where the level density is large, is very similar. As a consequence, the columns of $\mathbf{R}$ are nearly degenerate,
making the matrix inversion numerically unstable.



In 2007, Tain and Cano-Ott~\cite{Tain:07,Tain:07a} carried out a comprehensive investigation of iterative unfolding algorithms, including their systematic uncertainties, and demonstrated their suitability for TAGS data analysis. More recently, attention has turned toward the application of modern Machine Learning (\emph{ML}) techniques to address some limitations inherent to traditional unfolding methods. In this context, Dembski \textit{et al.}~\cite{Dembski:24} pioneered the case of multi-crystal spectrometers, introducing the use of conditional Generative Adversarial Networks to unfold two-dimensional (E$_{x}$, E$_{\gamma}$) matrices, formulating the problem as an image-to-image translation task. Their work provides a proof of concept, using both simulated and experimental data from simple decay schemes populating a single level and involving either a single $\gamma$-ray or a two-step cascade (e.g., $^{137}$Cs or $^{60}$Co). While successful, this approach was demonstrated only for low-complexity cases characterized by very low level densities and only a few distinct $\beta$ transitions~\cite{Dembski:24}.

In the present work, we restrict ourselves to one-dimensional problems, performing a systematic comparison between conventional Response-Matrix-based unfolding methods (\emph{RM}) and contemporary ML techniques (supervised \emph{ML}) for TAGS data analysis. Specifically, we benchmark three established response-matrix-based algorithms —Richardson–Lucy/Expectation Maximization (BAYES)~\cite{Richardson:72,Lucy:74,DAGOSTINI:95}, Differential Evolution (DIFEVO)~\cite{Storn:97}, and Non-Negative Least Squares (NNLS)~\cite{Lawson:95}— against supervised approaches based on Convolutional Neural Networks (CNN)~\cite{Bishop:24} and Gradient Boosted Decision Trees (GBDT)~\cite{Coadou:22}. Unlike the traditional methods, the supervised \emph{ML} models aim to determine the feeding vector $\vec{f}$ directly from the measured spectrum, without explicitly using the response matrix $\mathbf{R}$ during the inference stage.
The performance of all algorithms is evaluated using the decay of a synthetic nucleus $^{152}$Tb$^{\dagger}$, based on the actual $^{152}$Tb as explained later, characterized by high level density and complex cascade structures, where the TAGS technique becomes essential for a reliable determination of the $\beta$ intensity~\cite{Tain:07,Tain:07a,Algora:07}. This comprehensive comparison aims to quantify the advantages and limitations of supervised \emph{ML} approaches relative to well-established \emph{RM} unfolding techniques in the context of complex nuclear TAGS problems.

This work is organized as follows. In Sec.~\ref{Sec:Meth}, we describe the algorithms considered in this study and the methodology adopted for their implementation and evaluation. The results are presented in Sec.~\ref{sec:Results}: Sec.~\ref{sec:Tb} discusses the results obtained for decay of $^{152}$Tb$^{\dagger}$, while Sec.~\ref{sec:stat} examines the performance of the different methods as a function of decreasing statistical quality in the input TAS spectra. Finally, the main findings and conclusions are summarized in Sec.~\ref{sec:Summary}.

\section{Methodology}\label{Sec:Meth}

As stated in the introduction, in this work we consider two distinct statistical strategies for estimating the level feedings: supervised \emph{ML} and \emph{RM} methods. These approaches rely on different statistical estimators and algorithmic frameworks, which are briefly detailed in this section.

\begin{figure}[ht]
\centering
\includegraphics[width=1.0\columnwidth]{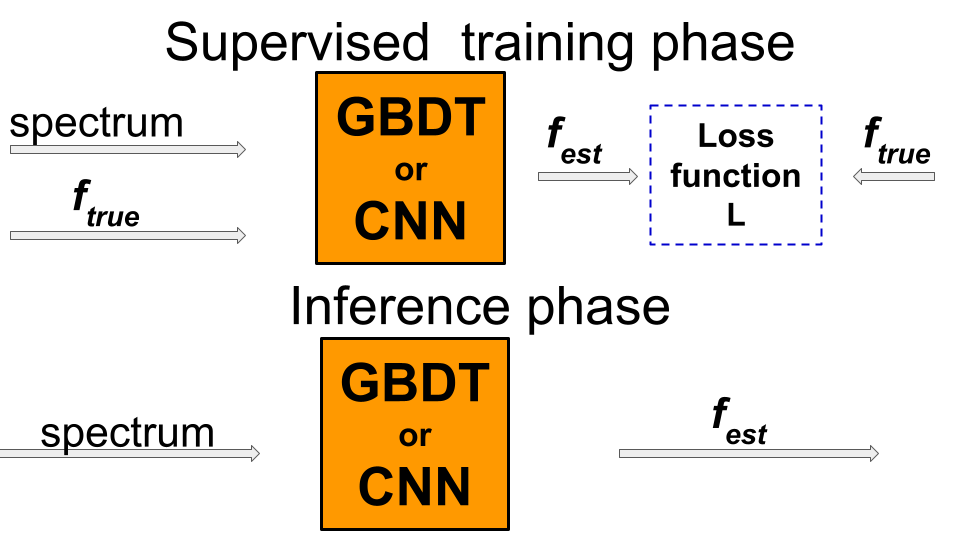}
\caption{Schematic representation of the supervised \emph{ML} estimator workflow. The top part illustrates the training phase, where the model learns from labeled data, while the bottom part depicts the inference phase, in which the trained model is applied to new input spectra to estimate the corresponding feeding distribution.}
\label{fig:Supervised_training}
\end{figure}

For supervised \emph{ML} level-feeding estimation, formulated as a supervised problem, a non-parametric statistical estimator of the feeding vector is constructed using a controlled dataset composed of paired spectra and corresponding true feeding distributions, \(\{\vec{d},\vec{f}_{\mathrm{true}}\}\) obtained either from experiments or more often from MC simulations. In this case, the estimator minimizes a suitable loss function that quantifies the discrepancy between the true feeding vector, \(\vec{f}_{\mathrm{true}}\), and the predicted one, \(\vec{f}_{\mathrm{est}}\)~\cite{bishop:07,murphy:13}. This optimization step, commonly referred to as the training phase, enables the supervised model to learn the mapping between the input spectrum \(\vec{d}\) and the feeding vector \(\vec{f}\). Once trained, the model enters the inference phase, in which it is applied to an experimental or MC spectrum to estimate the unknown feeding distribution, \(\vec{f}_{\mathrm{est}}\). The overall workflow of the supervised \emph{ML} training and inference procedure is illustrated in Fig.~\ref{fig:Supervised_training}.

In this work, we employed two algorithms for supervised \emph{ML} level-feeding estimation: CNN and GBDT, implemented using the \textsc{TensorFlow}~\cite{tf-wp:15} and \textsc{XGBoost}~\cite{xgboost:25} libraries, respectively.

\begin{figure}[ht]
\centering
\includegraphics[width=1.0\columnwidth]{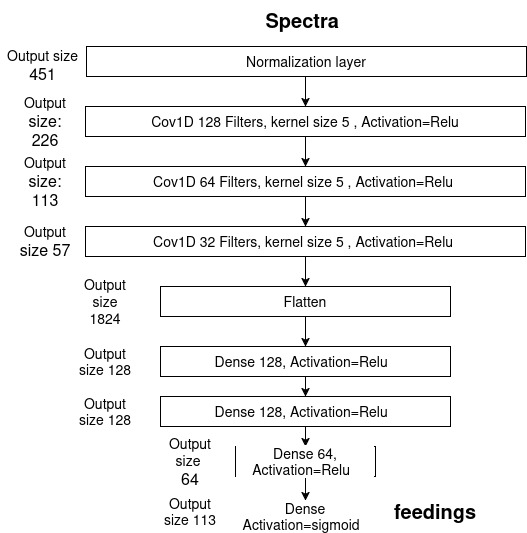}
\caption{Schematics of the CNN architecture employed in this work for \(^{152}\)Tb$^{\dagger}$. On the left the output dimension of each layer is indicated.}
\label{fig:NNArchitecture}
\end{figure}

The CNN architecture used in this work begins with an input normalization layer, followed by three one-dimensional convolutional layers with a kernel size of 5 and a stride of 2. These layers use 128, 64, and 32 filters, respectively, progressively extracting higher-level features from the input spectrum. The output of the final convolutional layer is flattened and fed into three fully connected layers comprising 128, 128, and 64 units. A rectified linear activation function is employed throughout the network to introduce nonlinearity, except in the output layer. The final output layer is a fully connected layer whose dimensionality is equal to the number of feeding components to be estimated, with a \textit{sigmoid} activation function that constrains the predicted feeding values to the interval \([0,1]\). A schematic representation of the network architecture, including the dimensionality of each layer is presented in Fig.~\ref{fig:NNArchitecture}. Two loss functions, \emph{L}$_{1}$ and \emph{L}$_{2}$, were considered during the CNN training process: the mean absolute error (MAE)
\begin{equation}
L_{1}=\frac{1}{N_{r}}\left|\vec{f}_{est} - \vec{f}_{true} \right|
\end{equation}

and the mean squared error (MSE)

\begin{equation}\label{Eq:MSE}
L_{2}=\frac{1}{N_{r}}\left|\left|\vec{f}_{est} - \vec{f}_{true} \right|\right|^{2}
\end{equation}

where $N_{r}$ is the dimension of $\vec{f}_{true}$, or equivalently, the number of levels involved in the unfolding. These two loss functions allow us to assess the sensitivity of the model performance to different error penalization schemes.

In contrast, GBDT is based on gradient boosting and combines decision-tree learners~\cite{Rokach:05} within an ensemble framework conceptually related to random forest methods~\cite{Breiman:01}, but optimized through a sequential boosting strategy~\cite{xgboost:25}. In this work, we employed an ensemble of 200 trees with a maximum depth of 5, a learning rate of 0.8, and a subsampling ratio of 80\% per tree, while keeping the remaining hyperparameters at their default settings. For this model, only the $L_{2}$ loss function stated in Eq.~\ref{Eq:MSE} was used as the optimization objective.

In the \emph{RM} estimation approaches, the statistical estimators are constructed to minimize a loss function that measures the discrepancy between the observed spectrum, \(\vec{d}\), and the reconstructed TAS spectrum, \(\vec{d}_{\mathrm{est}}\) from the estimated feeding vector, \(\vec{f}_{\mathrm{est}}\) (Eq.~\ref{eq:1}). In general, the algorithms are iterative starting from an initial guess or \emph{a priori} solution. Typically, the loss function incorporates, via a Lagrange multiplier, a regularization functional~\cite{Titterington:1985,Tikhonov:1963} on the solution \(\vec{f}_{\mathrm{est}}\) to counteract the ill-conditioned nature of the inverse problem and remove nonphysical oscillations on the solution. Sometimes the regularization condition is implicit in the form of the unfolding algorithm. A major advantage of these methods over the supervised \emph{ML} approaches is that they do not require a training phase, as the solution is obtained directly from the experimental data through the unfolding procedure.

We applied three different \emph{RM} algorithms within this framework: DIFEVO, BAYES, and NNLS, implemented using the \textsc{DEAP}~\cite{DEAP:12}, \textsc{PyUnfold}~\cite{pyUnfold:18}, and \textsc{SciPy}~\cite{SciPy:20} libraries, respectively. Each algorithm differs in the way the estimated feeding is updated in each iteration. The DIFEVO is a stochastic optimization algorithm based on population dynamics~\cite{Storn:97}. Unlike the other two algorithms, DIFEVO updates the solution
space through mutation and recombination operations applied to a population of candidate solutions, which makes it less sensitive to local minima and suitable for highly nonlinear problems. In our case, a constant population of 300 individuals was adopted, with a crossover probability of 0.9 and a mutation rate of 0.1. A hall-of-fame size of 30 and a crowding factor of 20 were used to preserve diversity within the population. The optimization was carried out over 800 generations. The components of feeding vector, \(\vec{f}_{\mathrm{est}}\), are constrained to lie within the interval \([0,1]\). The loss function to be minimized was the negative of the Poisson log-likelihood, \(l\), between the observed and reconstructed spectra, defined as

\begin{equation}\label{eq:l}
l=\sum_i \left(N\,\mathbf{R}\cdot\vec{f}_{\mathrm{est}}\right)_i
-\sum_i d_i\log\left[\left(N\,\mathbf{R}\cdot\vec{f}_{\mathrm{est}}\right)_i\right].
\end{equation}


BAYES is an algorithm based on the iterative application of Bayes' theorem~\cite{Bayes:1764}, implemented here following the Richardson--Lucy/Expectation-Maximization scheme~\cite{Richardson:72,Lucy:74,DAGOSTINI:95}, which relates causes and effects while requiring minimal \textit{a priori} information on the solution. In this work we considered a flat non-informative prior as starting point for the algorithm. The convergence criterion for the iterative procedure was defined using a minimum value threshold ($\varepsilon$) for the squared Euclidean distance between the estimated feeding vectors obtained in two consecutive iterations, $i$ and $i+1$,

\begin{equation}
\left|\left|\vec{f}_{est,i+1}-\vec{f}_{est,i}\right|\right|^{2}< \varepsilon
\end{equation}

Thus, the iterative procedure was terminated once the convergence criterion was satisfied ($\varepsilon=0.01$) or after a maximum of 200 iterations. To stabilize the solution and mitigate fluctuations, a third-order spline regularization was applied during the minimization process.

In the case of NNLS the minimization of the loss function 
\begin{equation}\label{eq:chi2}
\chi^{2} = \sum_i \left( d_i - \left(N\,\mathbf{R}\cdot\vec{f}_{\mathrm{est}}\right)_i \right)^2 / \sigma^{2}_{i}
\end{equation}
is performed under the additional constraint that all $\vec{f}_{est}$ of the solution remain non-negative using an active/passive sets method to solve the Karush-Kuhn-Tucker conditions~\cite{Lawson:95}. The non-negative condition acts effectively as a regularization condition. The per-bin uncertainty $\sigma_i$ was taken as the Poisson standard deviation, $\sigma_i=\sqrt{d_i}$, with bins having $d_i=0$ excluded from the sum to avoid a numerical divergence.

Both supervised \emph{ML} and \emph{RM} methodologies were investigated using MC datasets that provide a realistic representation of the TAGS problem, including the appropriate statistical fluctuations. For supervised approaches in particular, the training of any statistical estimator must rely on extensive representative
MC or experimental data. This constitutes a critical aspect of supervised \emph{ML} methods, since only these allow the model to learn accurate and physically meaningful solutions. The TAGS feeding-retrieval problem is no exception. Accordingly, two million MC-generated samples were used in the training and validation of all supervised estimators in the case studied in this work.

Each sample consisted of a pair comprising a TAS spectrum and its associated true feeding distribution, \(\{\vec{d},\vec{f}_{\mathrm{true}}\}\). The individual feeding components were generated by sampling around the central values of a chosen reference feeding vector \(\vec{f}_{\mathrm{ref}}\) assuming a flat probability distribution in a given interval. Each feeding component was varied independently up to a maximum of 10\% of the nominal value, and the resulting feeding vector was normalized such that
\[
\sum_i f_{\mathrm{true},i}=1.
\]
Negative feedings were set to 0, as they have no physical meaning. The corresponding TAS spectra for each MC sample were generated in histogram form, incorporating statistical fluctuations. Specifically, the content of each bin, \(d_i\), was sampled from a Poisson distribution:
\begin{equation}\label{eq:2}
    d_i \sim \mathrm{Poisson}\left(N\,(\mathbf{R}\cdot\vec{f}_{\mathrm{true}})_i\right).
\end{equation}

In the preceding equation, \(N\), as in previous cases, represents the total number of decays and thus defines the statistical accuracy of the simulated data. The introduction of Poisson-distributed fluctuations reproduces the intrinsic counting statistics expected in a TAGS experiment at any statistical level.

The TAS response function required for each individual level, was calculated by MC simulations using a detailed \textsc{GEANT4}~\cite{Allison:16} geometrical model of the Lucrecia detector currently installed at ISOLDE~\cite{Rubio:17}. The Electromagnetic Physics List Option 4 was employed together with the Radioactive Decay module to account for all relevant de-excitation processes, including \(\gamma\)-ray emission, but also X-ray production, and Auger-electron emission if the conversion electron process occurs. 
The resulting response functions were stored in histogram format using the \textsc{ROOT} framework~\cite{ROOT:97}. The histograms were defined with a fixed bin width of 10~keV.

The performance of all algorithms was assessed using an independent validation dataset consisting of 1000 spectrum--feeding pairs that were not used during the training of the supervised models. Each vector \(\vec{f}_{\mathrm{true}}\) in this sample was generated following the same procedure described previously in this section, ensuring consistency between the training and evaluation conditions.

The evaluation of the resulting \(\vec{f}_{\mathrm{est}}\) was first performed at the level of individual feeding components by computing the difference between the estimated and true values. This level-by-level analysis provides insight into the expected accuracy of each method and enables the identification of potential systematic biases affecting specific $\beta$-feedings. In a second step, a global assessment of the performance was conducted by examining the quality of the reconstructed spectra using the Baker-Cousins $\chi^2_{BC}$ statistic test~\cite{Baker:1984}, together with the $L_{2}$ metric which is an overall measure of the accuracy in estimating the full feeding vector. The Baker-Cousins statistical test is defined as

\begin{equation}\label{eq:chiBC}
\chi^2_{BC}=2\sum_i \left[\left(N\,\mathbf{R}\cdot\vec{f}_{\mathrm{est}}\right)_i - d_i + d_i\log\left(\frac{d_i}{\left(N\,\mathbf{R}\cdot\vec{f}_{\mathrm{est}}\right)_i}\right)\right],
\end{equation}

with the convention that the term $d_i\log(d_i/(N\,\mathbf{R}\cdot\vec{f}_{\mathrm{est}})_i)$ is set to zero for bins with $d_i=0$. This statistic provides a Poisson-consistent goodness-of-fit measure between the observed and reconstructed spectra, remaining well behaved even in low-count bins where a conventional least-squares $\chi^2$ is not applicable. For convenience, during the discussion of the results, the $\chi^2_{BC}$ was divided by the number of reconstructed spectra bins, $N_{b}$.

For the evaluation of the overall feeding reconstruction, we will use the $L_{2}$ metric defined in Eq~\ref{Eq:MSE} because of its robustness and easy interpretation since it represents the average square difference between the true and estimated feeding vectors. 

The evaluation of all algorithms was initially performed using spectra with very high statistics corresponding to an ideal TAGS experiment. Subsequently, the statistics of the input spectra were progressively reduced and the performance analyzed. This approach provides a comprehensive perspective not only on the intrinsic performance of each algorithm, but also on their robustness as the statistical quality of the data varies.

\section{Results and discussion}\label{sec:Results}

In this section, the performance of the different methodologies is evaluated for the $^{152}$Tb$^{\dagger}$ decay scenario described below. The results from the statistical evaluation are presented in Sec.~\ref{sec:Tb}. Finally, in Sec.~\ref{sec:stat}, we investigate the impact of reduced counting statistics in the initial spectra on the performance of the different algorithms.

\subsection{$^{152}$Tb$^{\dagger}$ decay results}\label{sec:Tb}

In this work, we consider the EC/$\beta+$ decay of $^{152}$Tb ($J^{\pi}=2^{-}$, $T_{1/2}=17.5$ h) to the daughter nucleus $^{152}$Gd, for which 113 levels are energetically accessible according to the Evaluated Nuclear Structure Data File (ENSDF)~\cite{Martin:13}. Starting from the ENSDF decay scheme, we construct a modified benchmark case, denoted as $^{152}$Tb$^{\dagger}$ and $^{152}$Gd$^{\dagger}$, by introducing a weakly fed continuum above $E_x = 2.9$ MeV. Although the highest populated level listed in ENSDF is located at $E_x = 3.458$ MeV, the continuum is assumed to start at lower excitation energy to account for the increasing level density and the corresponding presence of weakly populated states that are missing from the evaluated level scheme. The continuum is discretized into 40-keV bins, and the $\gamma$-ray de-excitation branching ratios are generated using the statistical-model prescription described in Ref.~\cite{Tain:07a}. The resulting discrete-plus-continuum level scheme extends up to the decay energy $Q_{EC}=3.99(4)$ MeV. The nominal $\beta$ intensities were taken from ENSDF for the discrete levels. For the continuum bins we adopted the corresponding ENSDF $\beta$ intensity, whenever available, and 0.1$\%$ when there was no information from ENSDF, to account for the weakly populated levels expected in this energy region. All spectra presented in this section were generated with a total statistics $N = 10^{8}$ decays, thereby representing an almost ideal TAGS scenario.

\begin{figure*}[htb!]
\centering
\includegraphics[width=2.0\columnwidth]{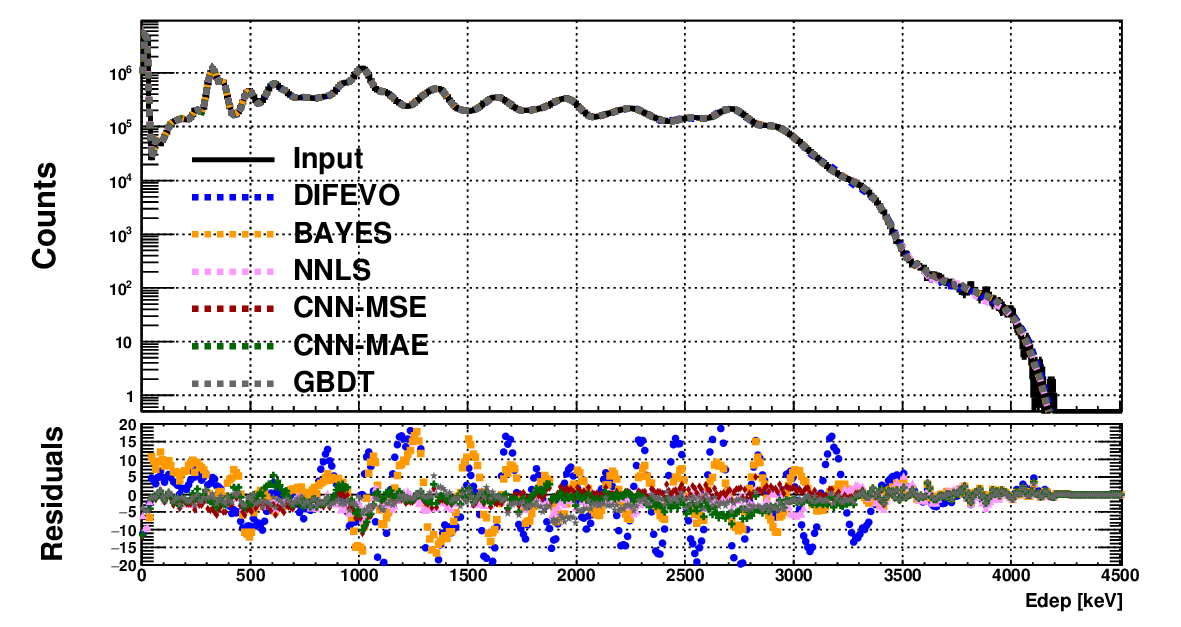}
\caption{Top: Input TAS spectrum (black) for $^{152}$Tb$^{\dagger}$ decay, obtained from the nominal feeding distribution, and reconstructed spectra (color) obtained from the estimated feedings for all methodologies considered in this work. 
Bottom: Calculated residuals between input and reconstructed TAS spectrum expressed in units of the statistical uncertainty. See the text for a detailed discussion.}
\label{fig:ResultsTb152_1}
\end{figure*}

\begin{table*}[htb!]
\centering
\begin{tabular}{|c|c|c|c|c|c|c|c|}
\hline
$E_{x}$ [keV] & nominal $f$ [\%] & DIFEVO & BAYES & NNLS & CNN-MSE & CNN-MAE & GBDT\\
\hline\hline
0.0 & 25.39 & 25.02 & 24.33 & 25.38 & 24.98 & 24.98 & 25.38\\ \hline
344.28 & 12.90 & 14.04 & 13.06 & 12.93 & 12.74 & 12.77 & 12.93\\ \hline
615.39 & 6.96 & 7.24 & 6.96 & 6.95 & 6.86 & 6.98 & 6.94\\ \hline
930.54 & 8.19 & 7.83 & 8.13 & 8.18 & 8.18 & 8.22 & 8.18\\ \hline
1047.8 & 1.37 & 2.18 & 2.20 & 1.32 & 1.37 & 1.37 & 1.32\\ \hline
1109.2 & 2.83 & 1.07 & 1.99 & 2.87 & 2.80 & 2.81 & 2.87\\ \hline
1123.19 & 1.70 & 3.02 & 2.39 & 1.63 & 1.72 & 1.72 & 1.63\\ \hline
1318.41 & 2.77 & 0.14 & 1.96 & 2.90 & 2.74 & 2.78 & 2.90\\ \hline
1605.61 & 2.33 & 1.90 & 2.07 & 2.36 & 2.29 & 2.27 & 2.36\\ \hline
1643.42 & 1.89 & 1.63 & 1.96 & 1.90 & 1.86 & 1.90 & 1.90\\ \hline
1861.58 & 1.00 & 0.04 & 0.33 & 0.99 & 0.98 & 1.02 & 0.99\\ \hline
1941.18 & 4.35 & 2.71 & 1.38 & 4.27 & 4.35 & 4.30 & 4.27\\ \hline
2246.81 & 3.81 & 0.33 & 1.60 & 3.76 & 3.77 & 3.78 & 3.76\\ \hline
2299.5 & 1.01 & 0.16 & 0.48 & 1.02 & 1.00 & 1.00 & 1.02\\ \hline
2709.42 & 1.68 & 1.00 & 0.48 & 1.68 & 1.66 & 1.63 & 1.68\\ \hline
2719.63 & 1.41 & 0.00 & 0.67 & 1.40 & 1.42 & 1.38 & 1.40\\ \hline
2729.17 & 1.02 & 1.07 & 0.43 & 1.02 & 0.99 & 1.01 & 1.02\\ \hline
2749.24 & 1.61 & 0.11 & 0.88 & 1.60 & 1.62 & 1.57 & 1.60\\ \hline
2880.67 & 1.85 & 0.78 & 0.40 & 1.87 & 1.89 & 1.80 & 1.87\\ \hline
2900 & 1.16 & 0.00 & 0.61 & 1.10 & 1.13 & 1.15 & 1.10\\ \hline
2980 & 1.08 & 1.06 & 0.89 & 1.08 & 1.08 & 1.06 & 1.08\\ \hline
E$_{x}>$2900 & 4.34 & 4.09 & 4.38 & 5.42 & 4.29 & 4.27 & 4.26 \\ \hline
\end{tabular}
\caption{Extracted feedings for the decay of $^{152}$Tb$^{\dagger}$ (only values greater than 1\% and sum of the continium contribution) obtained with all methodologies considered in this work. The table lists the excitation energy of each level, the corresponding nominal values, and the feedings reconstructed by each algorithm for direct comparison.}
\label{tab:Tb152}
\end{table*}

Figure~\ref{fig:ResultsTb152_1} presents an example of the spectra reconstructed with each of the methodologies evaluated in this work, using the same input spectrum corresponding to the nominal $^{152}$Tb$^{\dagger}$ $\beta$-intensity distribution. The bottom panel shows the residuals between the reconstructed and input spectra, expressed in units of the corresponding statistical uncertainty. A visual inspection of top panel indicates that all methods achieve an overall good agreement with the input spectrum. However, a more detailed examination of the residuals reveals differences in the reconstruction. In particular \emph{RM} methodologies: DIFEVO, BAYES and NNLS, produce an oscillatory pattern in the residuals as a function of the deposited energy with values as large as 10--15 $\sigma$ for the two first methods, and 5 $\sigma$ for the latter. These systematic patterns point to an intrinsic limitation of these methods arising from the finite energy resolution and the similarity of the detector response to neighboring levels. As a result, strong correlations and compensation effects develop between the reconstructed feedings of adjacent levels in order to preserve both the global and local counting statistics of the measured spectrum. This leads to oscillatory structures in the reconstructed feeding distribution (discussed below), which are subsequently reflected in the reconstructed spectrum. By contrast, the supervised \emph{ML} approaches studied in this work: CNN-MSE, CNN-MAE and GBDT yield more stable reconstructions under the same conditions, with residuals that are generally smaller than 5 $\sigma$, therefore indicative of {\it{a priori}} more accurate recovery of the underlying feeding distribution. 

The feeding estimates obtained with all reconstruction methods are summarized in Table~\ref{tab:Tb152}. For clarity, only levels with nominal feeding intensities exceeding 1\% and the total continuum contribution are reported. Overall, the supervised \emph{ML} methods (GBDT, CNN-MSE, and CNN-MAE) provide feeding estimates for the discrete levels that are generally closer to the nominal values than those obtained with the \emph{RM} methods. The main exception is NNLS, whose performance is comparable, still worse, to that of the supervised \emph{ML} approaches. In contrast, for the continuum contribution, BAYES method yields the estimate closest to the nominal value, whereas all other methods deviate by more than 5\%.

\begin{figure}[htb!]
\centering
\begin{tabular}{c}
\includegraphics[width=1.0\columnwidth]{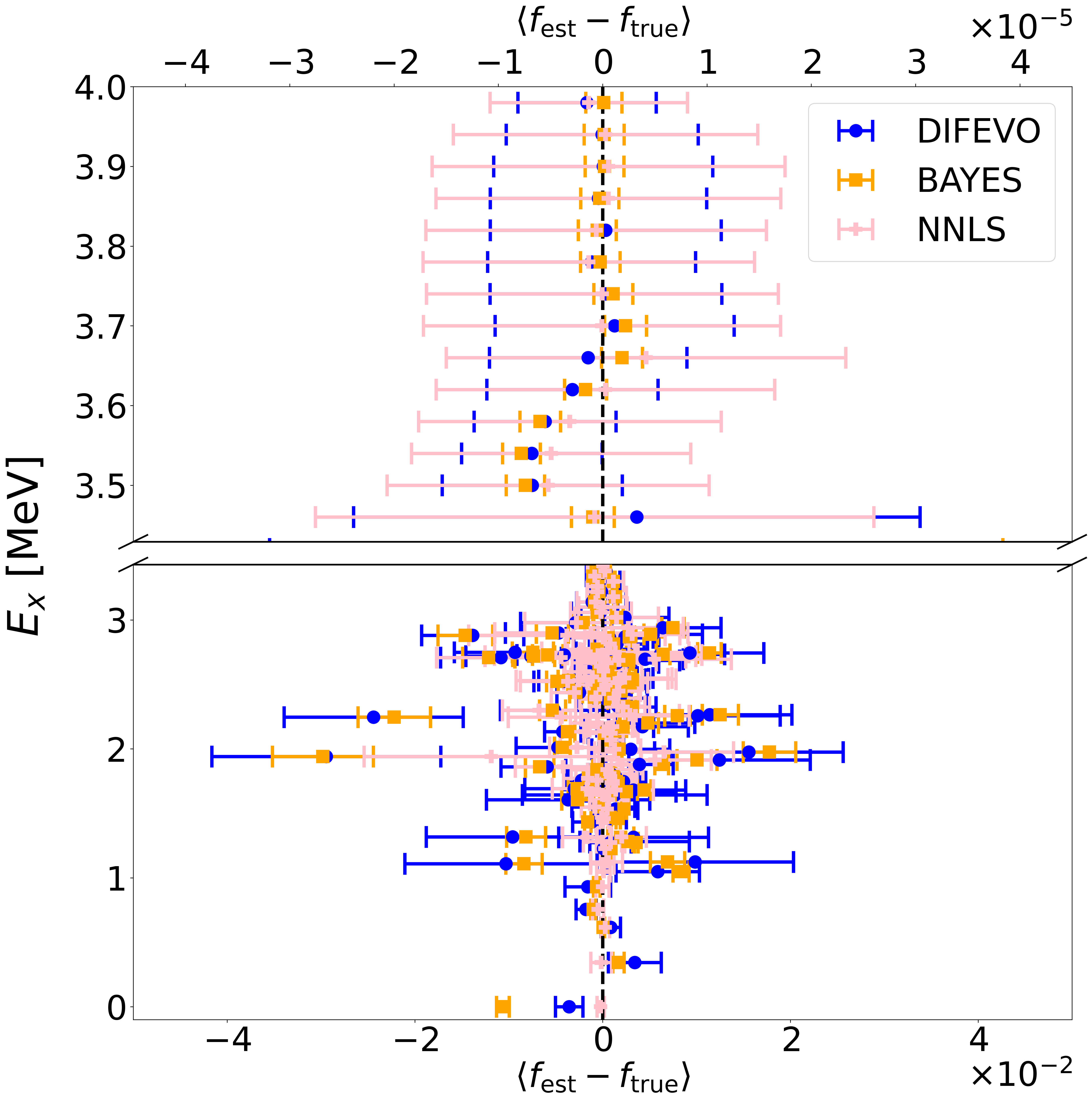} \\
\includegraphics[width=1.0\columnwidth]{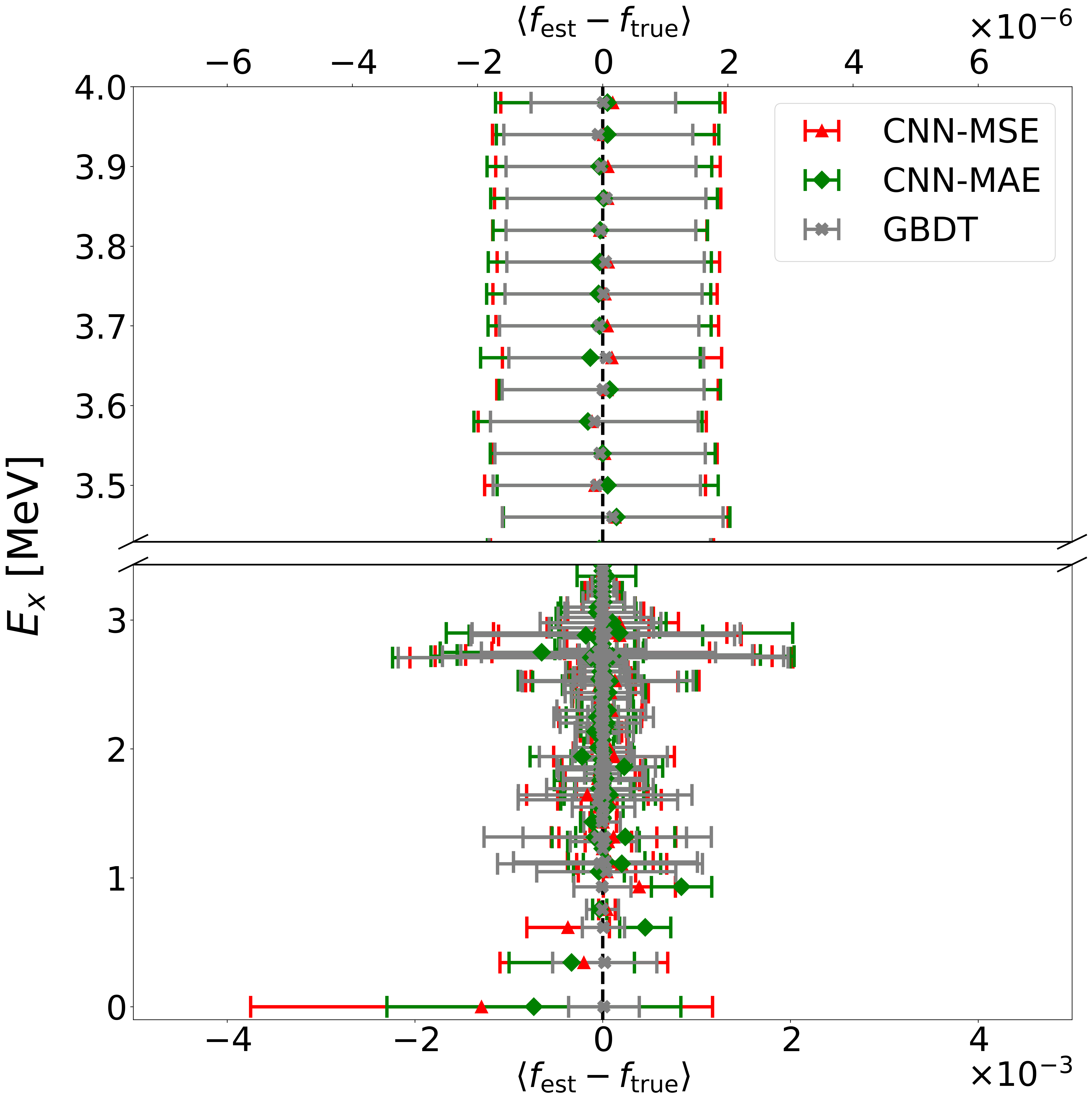}
\end{tabular}
\caption{Average difference between the estimated and true feedings for all individual levels $^{152}$Gd as a function of the excitation energy for all methodologies. Top: Results for \emph{RM} methodologies. Bottom: Results for supervised \emph{ML} methodologies. See the text for more details.}
\label{fig:ResultsTb152_2}
\end{figure}

Figure~\ref{fig:ResultsTb152_2} compares, for all excited levels, the estimated and nominal feeding intensities in the full validation dataset through the mean difference between the reconstructed and true values, together with the corresponding standard deviation. The top panel presents the results obtained with the \emph{RM} methods, whereas the bottom panel shows those from the supervised \emph{ML} methods. Because the feeding intensities and their associated uncertainties span a wide dynamic range, two horizontal scales are used in the top and bottom panels: the lower axis covers excitation energies from 0 to 3.2~MeV, while the upper axis displays levels above 3.2~MeV.

A clear difference is observed between the \emph{RM} and supervised \emph{ML} methods in both excitation-energy regions. For excitation energies below 3.2~MeV, the DIFEVO and BAYES methods exhibit a pronounced oscillatory pattern in the mean differences respect to 0, characterized by alternating positive and negative deviations of the order of 2-3 $\sigma$. This behavior is largely suppressed in the NNLS, CNN-MSE, CNN-MAE, and GBDT feeding reconstructions. An additional feature is also observed in the corresponding standard deviations: DIFEVO, BAYES, and NNLS yield uncertainties of the order of $10^{-2}$, whereas CNN-MSE, CNN-MAE, and GBDT reduce them by approximately one order of magnitude, to the $10^{-3}$ level. This corresponds to a factor 10 improvement in accuracy reconstructing the individual feedings for CNN-MSE, CNN-MAE, and GBDT methodologies.

A comparable behavior is found for excitation energies above 3.2~MeV, although the oscillatory pattern becomes substantially less pronounced for all reconstruction methods. In fact, all methods are compatible with zero within standard deviation values. However, the same hierarchy is observed reconstructing the standard deviation, with DIFEVO, BAYES, and NNLS achieving values of the order of $10^{-5}$, while CNN-MSE, CNN-MAE, and GBDT further reduce them to the $10^{-6}$ level. Overall, these results demonstrate the superior performance of the supervised \emph{ML} methods, particularly CNN-MSE, CNN-MAE, and GBDT, in reconstructing individual feeding intensities in high-complexity TAGS analyses. 

\begin{figure}[htb!]
\centering
\includegraphics[width=1.0\columnwidth]{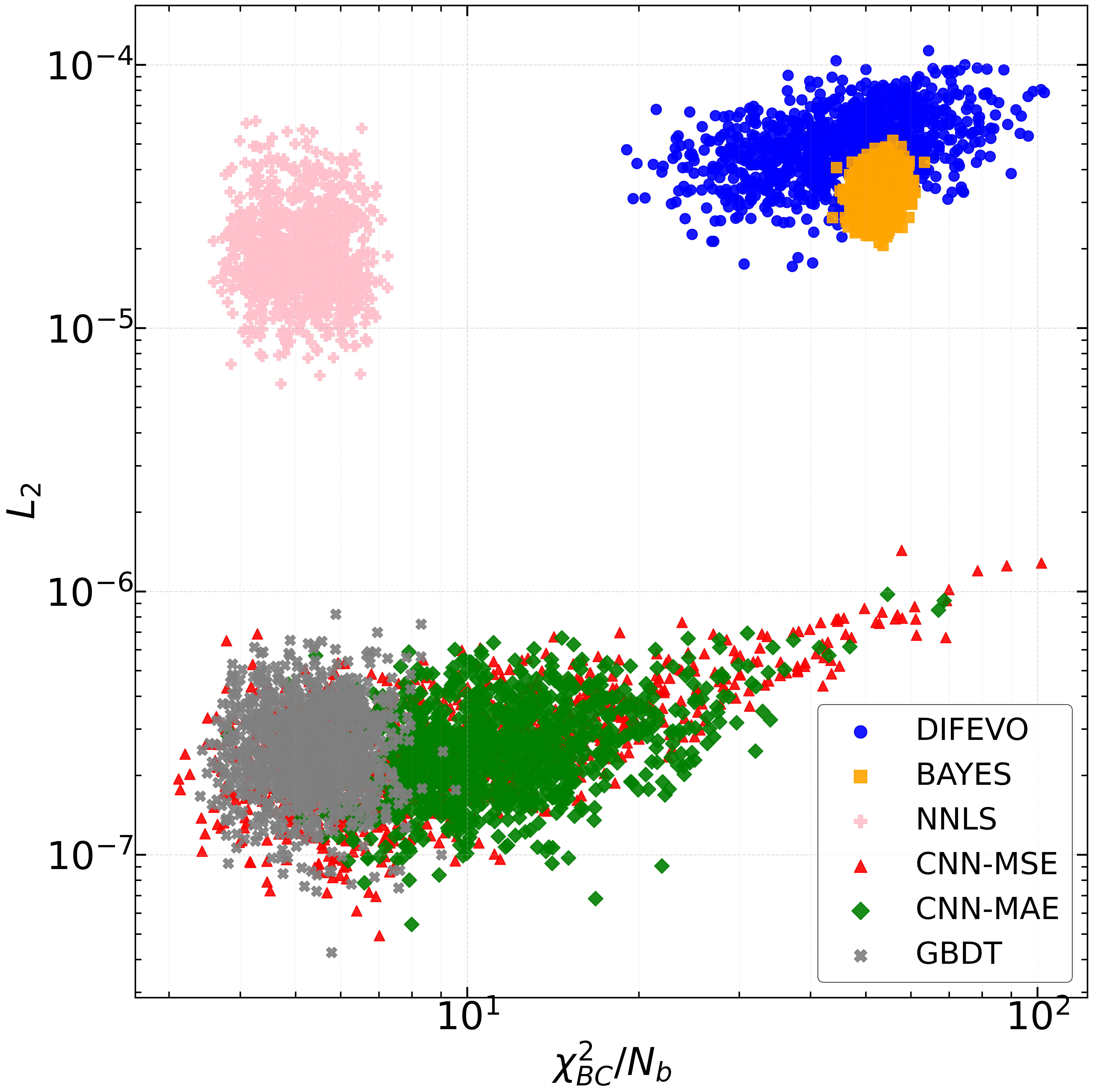}
\caption{Two-dimensional $\chi^2_{BC}$-$L_{2}$ scatter plot for $^{152}$Tb$^{\dagger}$ decay studied in this work. Each point corresponds to an individual TAS reconstruction from the test dataset, allowing a simultaneous evaluation of spectral reproduction and feeding recovery. See text for further details.}
\label{fig:ResultsTb152_3}
\end{figure}

As described in Sec.~\ref{Sec:Meth}, the overall reconstruction performance was assessed using the two-dimensional $\chi^2_{BC}/N_{b}$--$L_{2}$ distribution shown in Fig.~\ref{fig:ResultsTb152_3}, obtained from the results obtained for the complete test dataset. The different reconstruction methods form well-defined and clearly separated clusters, highlighting their distinct performance characteristics.

Among the \emph{RM} methods, DIFEVO and BAYES exhibit the poorest overall performance, with cluster centroids located around $L_{2}\sim5\times10^{-5}$ and $\chi^2_{BC}/N_{b}\sim50$. Although both methods achieve similar average reconstruction quality, the spread of the DIFEVO distribution is approximately three times larger than that of BAYES, indicating a significantly lower reconstruction stability. NNLS provides the best spectrum reconstruction among all methodologies, with $\chi^2_{BC}/N_{b}$ values centered around 4, while yielding intermediate feeding reconstruction accuracy with $L_{2}\approx2\times10^{-5}$. This comparatively low $\chi^2_{BC}/N_{b}$ suggests that NNLS reproduces the measured spectra particularly well, although this improvement does not translate into the most accurate recovery of the underlying feeding distribution, likely because of mild overfitting to the spectra data. Thus, capturing the statistical fluctuations observed in the spectra.

The supervised \emph{ML} methods, CNN-MAE, CNN-MSE, and GBDT, achieve the lowest $L_{2}$ values, with distributions centered around $2\times10^{-7}$, reflecting their excellent capability to reconstruct the feeding intensities. This behavior is expected, as these models are explicitly trained to minimize the discrepancy between the true and reconstructed feeding distributions. The three supervised ML approaches yield $\chi^2_{BC}/N_{b}$ values that are comparable to those obtained with NNLS. Among them, GBDT provides the most consistent spectrum reconstruction, as evidenced by the narrowest $\chi^2_{BC}/N_{b}$ distribution. In contrast, CNN-MSE and CNN-MAE achieve similar mean $\chi^2_{BC}/N_{b}$ values but display broader distributions with high-value tails extending up to $\chi^2_{BC}/N_{b}\approx80$, reflecting a limited number of reconstructions with noticeably poorer spectral agreement among the supervised \emph{ML} methods.

\subsection{Effect of input statistics on the results}\label{sec:stat}

We next investigate the impact of the input-spectrum statistics on the quality of the reconstructed $^{152}$Tb$^{\dagger}$ feeding distribution and spectra. To this end, the validation dataset described in Sec.~\ref{sec:Tb} was regenerated with progressively lower counting statistics, reducing the total number of decays from $N=10^{8}$ to $N=10^{4}$.

Since the supervised \emph{ML} models were trained exclusively with spectra containing $N=10^{8}$ counts, an additional preprocessing step was required for spectra with reduced statistics. First, spectra with $N=10^{4}$--$10^{7}$ counts were generated, preserving the Poisson fluctuations corresponding to each statistics level. They were then rescaled by a global normalization factor so that their total number of counts matched $N=10^{8}$. This procedure preserved the statistical fluctuations associated with each counting level while maintaining the input normalization expected by the trained supervised \emph{ML} models.

For each statistics level, the reconstruction performance was evaluated by calculating the mean values and standard deviations of the $L_{2}$ and $\chi^2_{BC}/N_{b}$ metrics over the corresponding validation dataset. This approach provides a quantitative assessment of both the average reconstruction quality and its statistical variability as a function of the available counting statistics.



\begin{figure}[htb!]
\centering
\begin{tabular}{c}
\includegraphics[width=1.0\columnwidth]{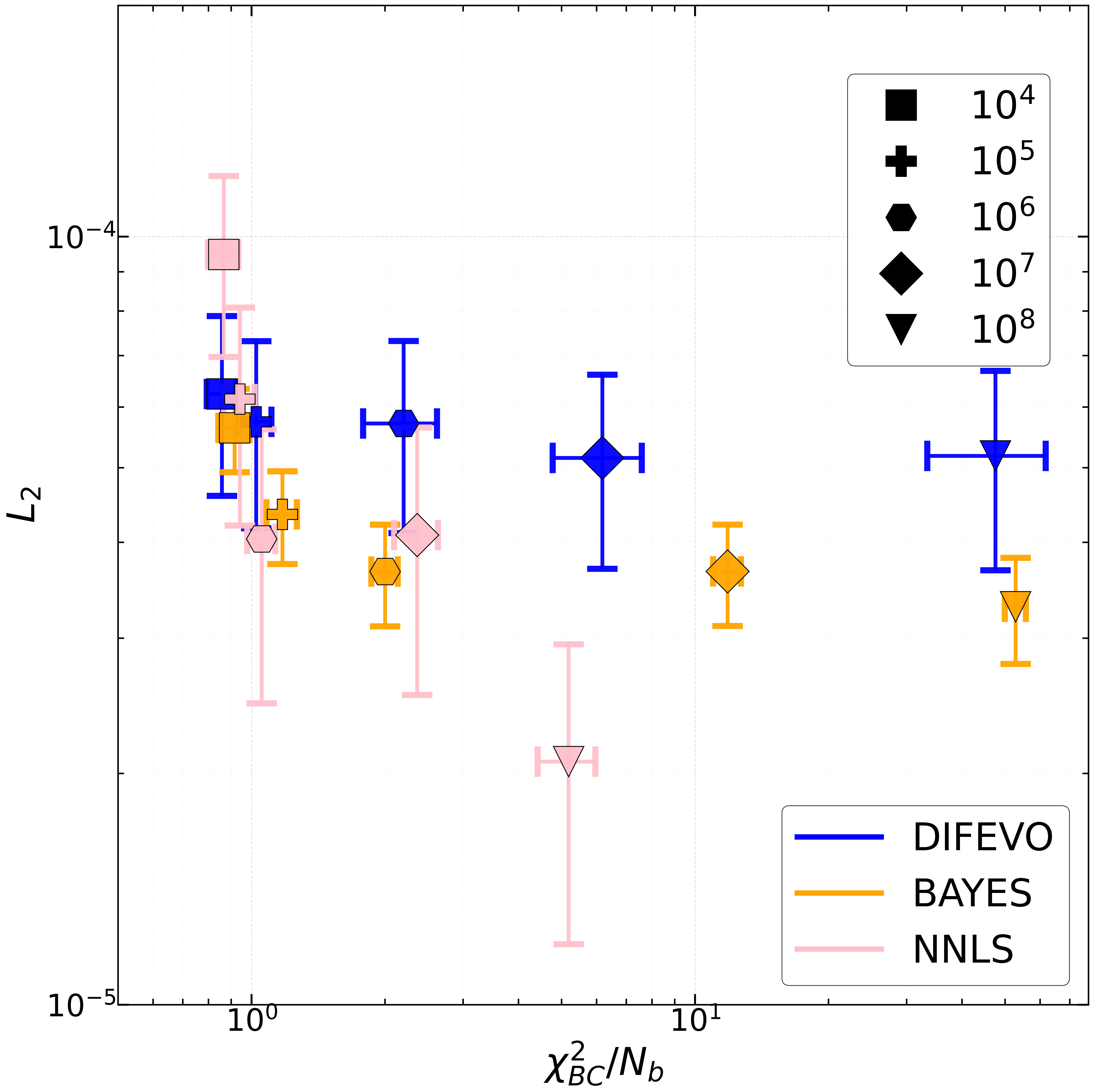}\\
\includegraphics[width=1.0\columnwidth]{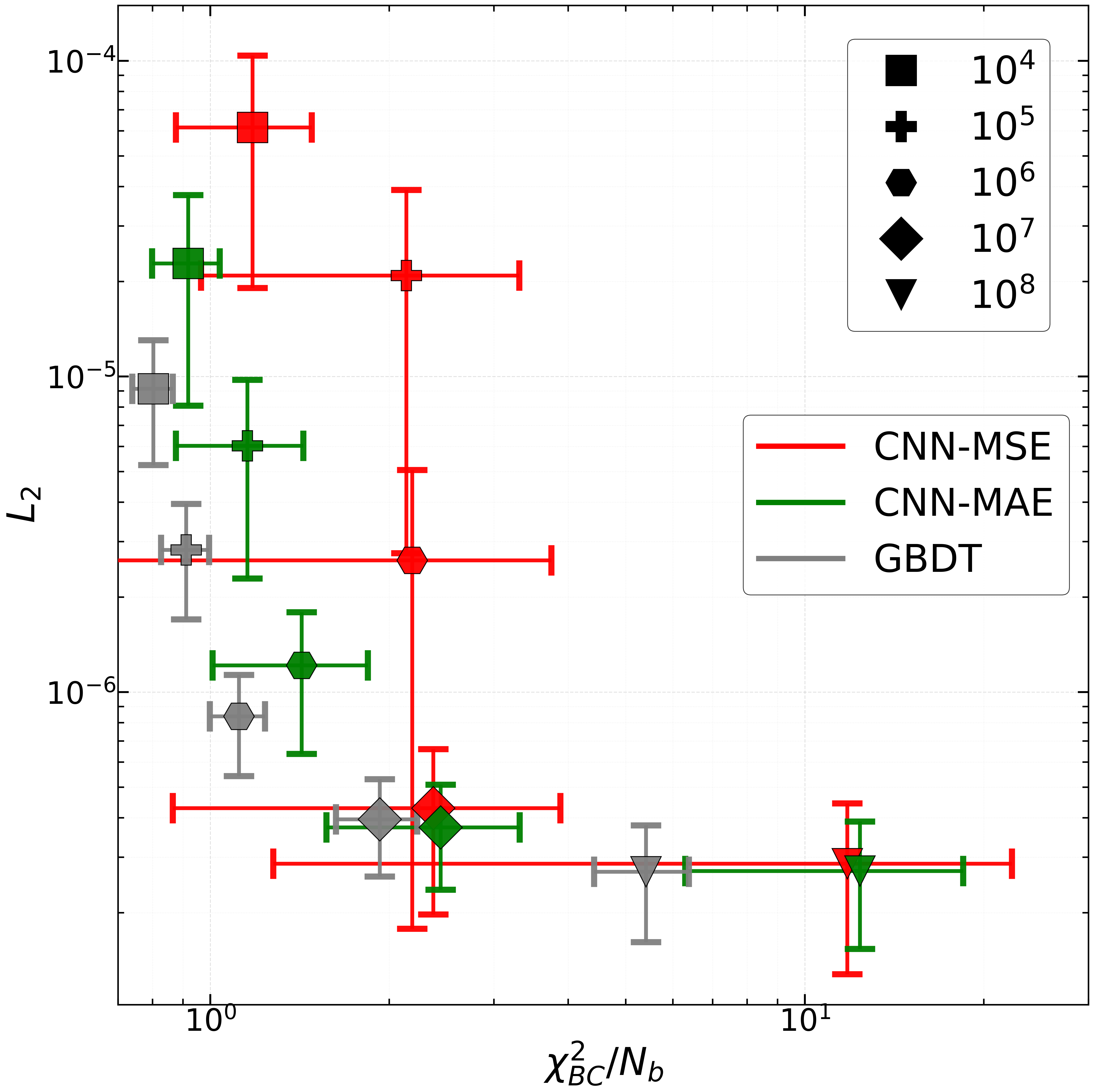}\\
\end{tabular}
\caption{Effect of input spectrum statistics on the $^{152}$Tb$^{\dagger}$ reconstruction performance. The horizontal axis shows the average value of $\langle\chi^2_{BC}/N_{b}\rangle$ , while the vertical axis displays the corresponding average $\langle L_{2}\rangle$. The error bars represent the standard deviation of each quantity over the test dataset. Top: Results for \emph{RM} methodologies. Bottom: Results for supervised \emph{ML} methodologies.}
\label{fig:Statistics_1}
\end{figure}

The dependence of both reconstruction performance on the total counting statistics is shown in Fig.~\ref{fig:Statistics_1}. For clarity, the top panel presents the \emph{RM} methods, whereas the bottom panel shows the supervised \emph{ML} methods. The different total statistics considered in this study, 10$^{4}$, 10$^{5}$, 10$^{6}$, 10$^{7}$, and 10$^{8}$ counts, are represented by squares, crosses, hexagons, diamonds, and inverted triangles, respectively. The color scheme adopted in Fig.~\ref{fig:ResultsTb152_1} is retained to facilitate comparison among the different reconstruction methods.

A markedly different dependence on the counting statistics is observed for the different methods studied in this work. Regarding feeding reconstruction among the \emph{RM} methods, DIFEVO and BAYES exhibit only a weak dependence on the available statistics, yielding nearly constant average $L_{2}$ values of approximately $4$--$5\times10^{-5}$ over the entire range considered. In contrast, NNLS shows a improvement with increasing statistics, reducing the average $L_{2}$ from approximately $\sim$10$^{-4}$ for 10$^{4}$ counts to about 2$\times$10$^{-5}$ for 10$^{8}$ counts.
At first sight, all \emph{RM} methods appear to exhibit a deterioration in the spectral reconstruction quality as the statistics increase, as reflected by the larger $\chi^2_{BC}/N_{b}$ values. For instance, DIFEVO, BAYES increases its average $\chi^2_{BC}/N_{b}$ from 1 up to 60 and 70, respectively. On the other hand, thanks to the better feeding reconstruction, NNLS reach a maximum of $\chi^2_{BC}/N_{b}$ close to four. This behavior, however, does not indicate a degradation of the reconstruction itself. Instead, it arises from the reduced statistical fluctuations of the input spectra at high counting statistics. At low statistics, the larger statistical uncertainties make a broader range of reconstructed spectra statistically compatible with the input data, naturally leading to smaller $\chi^2_{BC}/N_{b}$ values.

The supervised ML methods exhibit a substantially different behavior, with the feeding reconstruction accuracy strongly improving as the counting statistics increase up to $N=10^{7}$ counts. 
GBDT reduces the average $L_{2}$ from approximately 1$\times$10$^{-5}$ to 2$\times$10$^{-7}$, CNN-MAE from 2$\times$10$^{-5}$ to 2$\times$10$^{-7}$, and CNN-MSE from 8$\times$10$^{-5}$ to 2$\times$10$^{-7}$. Overall, all three supervised \emph{ML} methods improve their feeding reconstruction accuracy by roughly two orders of magnitude over the investigated statistics range. A conclusion similar to the \emph{RM} case can be extracted respect to the input spectra reproduction, with larger $\chi^2_{BC}/N_{b}$ for larger input statistics. All methodologies are close to 1 values for low statistics while in this case the maximum of the mean corresponds to CNN-MSE and CNN-MAE with values approximately 10. Thus, the same effect ascribed to the statistical fluctuations can be applied to supervised \emph{ML} methodologies.

In summary, although all methodologies 
exhibit a similar behavior in terms of $\chi^2_{BC}/N_{b}$ as function of input statistics, only the supervised \emph{ML} approaches are able to translate the reduction of statistical fluctuations in the input data into a significantly more accurate reconstruction of the feeding distributions. By contrast, \emph{RM} DIFEVO and BAYES methods, yield feeding estimates that remain essentially unchanged over the full range of statistics considered. Finally, NNLS shows a slight improvement in performance with increasing statistics, although the gain is small compared with that observed for supervised \emph{ML} methods. This behavior among \emph{RM} methods suggests that, although these approaches are robust, they have intrinsic methodological limitations that restrict their ability to benefit from improved statistical quality.

\section{Summary and conclusions}\label{sec:Summary}
Reliable extraction of $\beta$-feeding distributions is essential for exploiting the TAGS technique in nuclei characterized by large $Q$-values and high level densities. Although the relationship between the measured spectrum and the feeding distribution, Eq.~\ref{eq:1}, is formally simple, its inversion defines an ill-posed and ill-conditioned linear inverse problem whose solution is intrinsically sensitive to statistical fluctuations and systematic uncertainties. The choice of deconvolution methodology is therefore not a mere technical detail: it directly determines the reliability of the physical quantities extracted from a TAGS measurement.

In this work, we have carried out a systematic, quantitative comparison between supervised \emph{ML} estimators (CNN-MSE, CNN-MAE, GBDT) and established \emph{RM} unfolding algorithms (DIFEVO, BAYES, NNLS), applied to a common, high-statistics ($10^{8}$-count) MC dataset comprising pair of \(\{\vec{d},\vec{f}_{\mathrm{true}}\}\). The supervised \emph{ML} approaches construct a non-parametric estimator that maps a measured spectrum directly onto a feeding vector after a dedicated training stage in which the discrepancy between predicted and true feedings is minimized. The \emph{RM} approaches instead infer the feeding distribution by minimizing the difference between measured and reconstructed spectra, without an explicit training phase. Both classes of methods were benchmarked on the decay of $^{152}$Tb$^{\dagger}$, chosen deliberately as a high level-density scenario in which the density of states is comparable to or exceeds the detector resolution -- precisely the regime in which unfolding is most difficult and most needed. Establishing the generality of the trends reported here across decays of different level densities and cascade complexities is left for future work.

The central finding of this work is that a good spectral fit does not guarantee an accurate feeding reconstruction. All algorithms visually reproduce the input spectrum (Fig.~\ref{fig:ResultsTb152_1}), yet the quantitative Baker-Cousins $\chi^2_{BC}$ assessment (Fig.~\ref{fig:ResultsTb152_3}) reveals substantial differences in spectral-reconstruction quality that visual inspection alone fails to expose, with NNLS and GBDT achieving the closest statistical agreement and DIFEVO and BAYES the poorest. Crucially, this $\chi^2_{BC}/N_{b}$ ranking does not track the accuracy of the recovered feeding distribution. This decoupling follows directly from the differing optimization targets of the two approaches: \emph{RM} methods minimize the residual between measured and reconstructed spectra with no explicit constraint on the feeding vector itself, so an accurate spectral reproduction need not imply an accurate feeding reconstruction; supervised \emph{ML} methods, by construction, minimize the error on the feeding vector directly, and it is precisely this difference in optimization target that drives their superior performance. The converse relation always holds: a correctly determined feeding distribution reproduces the input spectrum by construction.

Quantitatively, the supervised \emph{ML} methodologies are statistically compatible with one another and consistently outperform the \emph{RM} methods across the full excitation-energy range (Fig.~\ref{fig:ResultsTb152_2}), reducing the uncertainty on individual feeding values by up to an order of magnitude. The \emph{RM} DIFEVO and BAYES methods, in contrast, display an oscillatory bias pattern in the reconstructed feedings that is intrinsic to the mathematical structure of these unfolding procedures rather than a statistical fluctuation. This bias both degrades their global feeding-reconstruction accuracy and drives the markedly larger $\chi^2_{BC}/N_{b}$ values obtained, exposing a formally poor spectral fit that remains invisible on visual inspection alone (Fig.~\ref{fig:ResultsTb152_3}). Interestingly, NNLS appears to suppress the oscillatory behavior, yielding values approximately three times smaller than those obtained with DIFEVO and BAYES. Among the supervised \emph{ML} methods, GBDT achieves the best spectra reproduction, outperforming, in average, CNN-MAE and CNN-MSE by approximately a factor of eight in $\chi^2_{BC}/N_b$.

On the $L_{2}$ metric, all three supervised \emph{ML} methods outperform the \emph{RM} methods by more than one order of magnitude, establishing their clear superiority for the most demanding requirement of TAGS analysis: the accurate recovery of individual feeding intensities.

The two families of methods respond in different ways to the counting statistics of the input spectrum (Fig.~\ref{fig:Statistics_1}). \emph{RM} methods are essentially insensitive to the available statistics in terms of feeding-reconstruction accuracy $\langle L_{2}\rangle$, while their $\chi^2_{BC}/N_{b}$ grows by more than an order of magnitude between $N=10^{4}$ and $N=10^{8}$: as statistical fluctuations shrink, the persistent oscillatory bias becomes an increasingly significant, statistically resolved contribution to $\chi^2_{BC}/N_{b}$. This comes at the cost of an inability to exploit improved statistical quality, and of an increasingly poor formal goodness-of-fit as statistics improve. Supervised \emph{ML} methods behave in the opposite way: their feeding-reconstruction accuracy improves steadily and substantially with increasing statistics, approaching their best performance in the high-statistics regime on which they were trained. Their superior accuracy comes at a price:
supervised \emph{ML} methods require good prior knowledge of the solution to the problem and a sufficiently large representative training dataset to reach optimal performance, whereas \emph{RM} methods require neither.

These complementary strengths motivate the central practical recommendation of this work: neither family of methods should be used in isolation. \emph{RM} methods remain indispensable for producing a physically consistent, training-free initial estimate of the feeding distribution -- an estimate that can, in turn, be used to construct the reference feeding vectors needed to train a supervised \emph{ML} model for the specific decay under study. The resulting \emph{ML} model then delivers a feeding reconstruction substantially more accurate than any \emph{RM} method achieves alone. We therefore propose a hybrid strategy for TAGS analysis, in which an \emph{RM} method first provides an initial feeding estimate that is subsequently refined by a supervised \emph{ML} model trained around it. This combined approach unites the robustness and training-free applicability of traditional unfolding with the superior accuracy of modern supervised methods, offering a practical route to more reliable feeding determinations precisely in the high level-density regime where existing unfolding methods are least reliable.

Two extensions of the present framework are intentionally deferred to a forthcoming publication: a systematic evaluation of the uncertainties associated with the reconstructed feeding distributions, which can be estimated from the dispersion observed over the ensemble of 1000 independent validation cases considered here, and the treatment of contaminants, environmental background, and pileup effects, which can be incorporated naturally as additional response components whose normalization factors are determined self-consistently within the deconvolution procedure.

\section*{Declaration of competing interest}
The authors declare that they have no known competing financial interests or personal relationships that could have appeared to influence the work reported in this paper.

\section*{CRediT authorship contribution statement}
\textbf{J. Balibrea-Correa:} Conceptualization, Investigation, Methodology, Supervision, Formal analysis, Data curation, Visualization, Writing - original draft.

\textbf{E. Nacher:} Conceptualization, Investigation, Methodology, Visualization, Writing -review \& editing.

\textbf{C. Fonseca-Vargas:} Investigation, Data curation, Visualization.

\textbf{J. L. Tain:} Investigation, Methodology, Visualization, Writing -review \& editing.

\section*{Data availability statement}
Data will be available on reasonable request.

\section*{Acknowledgments}
The authors acknowledge support from all the funding agencies of participating institutions. Part of this work was supported by the MCIN/AEI 10.13039/\-501100011033 under grants Severo Ochoa CEX\-2023-001\-292\--S, PID2022-138297NB-C21 and Generalitat Valenciana under CIPROM/2021/064 grant. 

\bibliographystyle{elsarticle-num} 
\bibliography{bibliography}






\end{document}